\documentclass[aps,prd,showpacs,showkeys,amsmath,amssymb]{revtex4}
\usepackage{dcolumn}
\usepackage{graphicx}
\usepackage{bm}
\usepackage{amsfonts}
\usepackage{amsmath}
\usepackage{latexsym}
\usepackage{amssymb}

\begin{document}

\title{The Possible Heavy Tetraquarks $qQ\bar q \bar Q$, $qq\bar Q \bar Q$ and $qQ\bar Q \bar Q$}
\author{Ying Cui, Xiao-Lin Chen}
\affiliation{Department of Physics, Peking University, Beijing
100871, China}
\author{Wei-Zhen Deng}
\email{dwz@th.phy.pku.edu.cn}\affiliation{Department of Physics,
Peking University, Beijing 100871, China}
\author{Shi-Lin Zhu}
\email{zhusl@th.phy.pku.edu.cn} \affiliation{Department of
Physics, Peking University, Beijing 100871, China}
\date{\today}
\begin{abstract}
Assuming X(3872) is a $qc \bar q \bar c$ tetraquark and using its
mass as input, we perform a schematic study of the masses of
possible heavy tetraquarks using the color-magnetic interaction
with the flavor symmetry breaking corrections.
\end{abstract}

\maketitle

\pagenumbering{arabic}

\section{Introduction}
\label{sec1}

The past several years have witnessed a renaissance of charmonium
spectroscopy. A few charmonium-like new resonances X(3872)
\cite{belle11}, X(3940) \cite{abe2}, Y(3940) \cite{belle12} and
Y(4260) \cite{aubert} were discovered experimentally. Because of
its very narrow width, proximity to $D{\bar D}^\ast$ threshold and
special decay patterns, Swanson proposed that X(3872) is mainly a
$D^0{\bar D}^{0\ast}$ molecular state with some admixture of $\rho
J/\psi$ and $\omega J/\psi$ components \cite{swanson1,swanson2}.
But the possibility of X(3872) as a conventional $1^{++}$
charmonium is still not excluded although its mass is smaller than
the quark model prediction \cite{barnes}. Li suggested that
X(3872) could be a charmonium hybrid meson \cite{li}. However,
both lattice QCD simulation \cite{liao} and flux tube mode
\cite{swanson} indicated that the hybrid charmonium lies in the
range 4200-4400 MeV.

Maiani et al. argued that the X(3872) is a bound state of a pair
of diquark and anti-diquark \cite{maiani}. With the spin-spin
interaction and X(3872) mass as input, they predicted a $2^{+}$
$cq \bar c \bar q$ tetraquark around 3952 MeV, which might be
identified to the X(3940) \cite{maiani}. This $2^{+}$ tetraquark
state can decay to $J/\psi$ plus a light vector meson, and
$D\bar{D}$ via D-wave.

The multi-quark system was studied many years ago using a quark model
with color-magnetic (CM) interaction \cite{belle1,belle2,chao1,chao2}.
Recently, Richard and Stancu used the model to estimate masses of $qc
\bar q \bar c$ tetraquarks \cite{richard1,richard2,richard3}. In the
complete $qc\bar q \bar c$ configuration space, they diagonalized the
$6\times 6$ Hamiltonian matrix. The CM interaction parameter was
determined by the charmed baryon masses. They found the mass of the
isoscalar state is around 3910 MeV, which lies well above X(3872) and
$D\bar {D^{*}}$ threshold.

In a previous work on light tetraquarks \cite{zhu-prd}, we found
that the strength of CM interaction in tetraquarks is quite
different from that in conventional $q\bar q$ mesons. The strength
of CM interaction in charmed baryons is also different from that
in the tetraquark system because tetraquarks and charmed baryons
have different spatial size. Hence the extraction of the CM
interaction parameter in the tetraquark system from the charmed
baryon is not justified.

In this work, we identify X(3872) as a tetraquark and use its mass
as input to fix the CM strength in the $cq\bar c\bar q$ system. In
Section \ref{sec2} we present the simple model Hamiltonian with
the color-magnetic interaction for the tetraquark system and
construct the color-spin-flavor wave functions of $0^+$, $1^+$ and
$2^+$ tetraquarks. Section \ref{sec3} is the numerical analysis.
The last section is a short summary.

\section{Heavy Tetraquark Masses}
\label{sec2}

For the tetraquark, the model Hamiltonian reads
\begin{equation}\label{p1}
H=\sum\limits_{i} m_i+H_{CM}
\end{equation}
where $m_i$ is the $i$-th constituent quark mass. $H_{CM}$ is the
color-spin interaction Hamiltonian which is derived from one-gluon
exchange in MIT bag model \cite{belle1,belle2,belle3,belle5}:
\begin{equation}\label{hcm}
H_{CM}=-\sum\limits_{i>j} v_{ij} \vec {\lambda_{i}} \cdot\vec
{\lambda_{j}}\vec {\sigma_{i}} \cdot\vec {\sigma_{j}}
\end{equation}
where ${\vec \sigma}_i$ is the quark spin operator and ${\vec
  \lambda}_i$ the color operator. For the anti-quark, ${\vec
  \lambda}_{\bar q}= -{\vec \lambda}^{*}$ and ${\vec \sigma}_{\bar
  q}=-{\vec \sigma}^{*}$. The values of the coefficients $v_{ij}$
depend on the multiquark system and specific models. In general,
$v_{ij}$ takes different values for $q\bar q$, $qqq$ and $q\bar
qq\bar q$ systems \cite{zhu-prd}. Here we discuss the heavy
tetraquarks, the $v_{ij}$ in the heavy quark system are also
different from the light quark system. We use the following
convention
\begin{equation}
\label{v-Q}
v_{ij}={\bar v}\frac{m_u^2}{m_i m_j}
\end{equation}
where ${\bar v}$ depends on the wave function of the multiquark
system. With given flavor context $q_1q_2\bar{q}_3\bar{q}_4$, the
expression of the CM matrix element between two states $|k\rangle$
and $|l\rangle$ reads
\begin{equation}\label{a0}
 V_{CM}(q_1q_2\bar q_3\bar q_4)=V_{12}(q_{1}q_{2})+V_{13}(q_{1}\bar{q}_{3})
 +V_{14}(q_{1}\bar{q}_{4})+V_{23}(q_{2}\bar{q}_{3})
 +V_{24}(q_{2}\bar{q}_{4})+V_{34}(\bar{q}_{3}\bar{q}_{4})
 \;,
\end{equation}
where $V_{ij}=\langle k \mid \vec{\lambda_{i}}\vec{ \lambda_{j}}
\vec{\sigma_{i}} \vec{\sigma_{j}} \mid l\rangle$. The base states
$|k\rangle$ and $|l\rangle$ are chosen to be the eigenstates of
$SU(6)_{cs}$.  Since the $|k\rangle$ and $|l\rangle$ are not
eigenstates of the CM interaction, we need calculate every
individual term in Eq.~(\ref{a0}) and diagonalize the $\langle k
\mid V_{CM} \mid l\rangle$ matrix in the base space in order to
obtain the physics mass.

We use the $qq\otimes \bar q \bar q$ to construct the color-spin
wave function of $0^+$, $1^+$ and $2^+$ heavy tetraquarks. A
particular multiquark configuration is denoted as
$|D_{6},D_{3c},S,N\rangle$, where $D_{6}$ and $D_{3c}$ are
$SU(6)_{cs}$ color-spin and $SU(3)_{c}$ color representations of
the multiquark system respectively. S is the spin of the system
and N is the total number of quarks and antiquarks.  The
$SU(6)_{cs}$ eigenstates of the $0^{+}$, $1^{+}$ and $2^{+}$ heavy
tetraquark are listed below:
\begin{eqnarray}
\label{en0-1}
&&|1,1_{c},0,4\rangle=\sqrt{\frac{6}{7}}|21,6_{c},1,2\rangle\otimes|\overline{21},\overline{6}_{c},1,2\rangle
+\sqrt{\frac{1}{7}}|21,\overline{3}_{c},0,2\rangle\otimes|\overline{21},3_{c},0,2\rangle\\
\label{en0-2}
&&|405,1_{c},0,4\rangle=\sqrt{\frac{1}{7}}|21,6_{c},1,2\rangle\otimes|\overline{21},\overline{6}_{c},1,2\rangle
-\sqrt{\frac{6}{7}}|21,\overline{3}_{c},0,2\rangle\otimes|\overline{21},3_{c},0,2\rangle\\
\label{en0-3} &&|1,1_{c},0,4\rangle=\sqrt{\frac{3}{5}}|15,
\overline{3}_{c},1,2\rangle\otimes|\overline{15},3,1,2\rangle
+\sqrt{\frac{2}{5}}|15,6_{c},0,2\rangle\otimes|\overline{15},\overline{6}_{c},0,2\rangle\\
\label{en0-4} &&|189,1_{c},0,4\rangle=\sqrt{\frac{2}{5}}|15,
\overline{3}_{c},1,2\rangle\otimes|\overline{15},3,1,2,\rangle
-\sqrt{\frac{3}{5}}|15,6_{c},0,2\rangle\otimes|\overline{15},\overline{6}_{c},0,2\rangle\\
\label{en1-1} &&|35,1_{c},1,4,\rangle =
|21,6_{c},1,2,\rangle\otimes|\overline{21},\overline{6}_{c},1,2,\rangle
\\
\label{en1-2} &&|35,1_{c},1,4\rangle =
|15,\overline{3}_{c},1,2,6_{f}\rangle\otimes|\overline{15},3_{c},1,2\rangle
\\
\label{en1-3} &&|35,1_{c},1,4\rangle =
\sqrt{\frac{1}{3}}|21,\overline{3}_{c},0,2\rangle
\otimes|\overline{15},3_{c},1,2\rangle
-\sqrt{\frac{2}{3}}|21,6_{c},1,2\rangle
\otimes|\overline{15},\overline{6}_{c},0,2\rangle
\\
\label{en1-4} &&|280,1_{c},1,4\rangle =
\sqrt{\frac{2}{3}}|21,\overline{3}_{c},0,2\rangle
\otimes|\overline{15},3_{c},1,2,\rangle
+\sqrt{\frac{1}{3}}|21,6_{c},1,2,\rangle
\otimes|\overline{15},\overline{6}_{c},0,2,\rangle
\\
\label{en1-5} &&|35,1_{c},1,4\rangle = \sqrt{\frac{1}{3}}
|15,\overline{3}_{c},1,2\rangle \otimes
|\overline{21},3_{c},0,2\rangle -\sqrt{\frac{2}{3}}
|15,6_{c},0,2\rangle \otimes
|\overline{21},\overline{6}_{c},1,2\rangle
\\
\label{en1-6}
 &&|280,1_{c},1,4\rangle =
\sqrt{\frac{2}{3}} |15,\overline{3}_{c},1,2\rangle \otimes
|\overline{21},3_{c},0,2\rangle +\sqrt{\frac{1}{3}}
|15,6_{c},0,2\rangle \otimes
|\overline{21},\overline{6}_{c},1,2\rangle
\\
\label{en2-1} &&|405,1_{c},2,4\rangle =|21,6,1,2 \rangle \otimes |
\overline{21},\bar
6,1,2 \rangle\\
\label{en2-2} &&|189,1_{c},2,4\rangle =|15,\bar 3,1,2 \rangle
\otimes | \overline{15},3,1,2 \rangle\;
\end{eqnarray}
Here we focus on the $0^+$, $1^+$ and $2^+$ states without radial
and orbital excitations. Their spatial wave functions are
symmetric. Therefore their color-spin wave functions of
flavor-symmetric diquark $\{qq\}$ or $\{\bar q\bar q\}$ inside
tetraquarks are antisymmetric. Those flavor-antisymmetric diquarks
$[qq]$ or $[\bar q \bar q]$ have symmetric color-spin wave
functions. We do {\sl not} assume that quarks in the tetraquark
really form diquarks with strong correlation. Instead we simply
use the $qq\otimes\bar q \bar q$ to construct the convenient base
wave functions.  On the other hand, one can also start from the
$q\bar q\otimes q \bar q$ basis. In order to calculate the
$V_{CM}(q\bar q)$, we must do some recoupling from the $|qq,
D_{6},D_{3c},S,2 \rangle \otimes | \bar q \bar q, D_{6},D_{3c},S,2
\rangle$ basis to the $|q \bar q, D_{6},D_{3c},S,2 \rangle \otimes
| q \bar q, D_{6},D_{3c},S,2 \rangle$ basis. Details can be found
in Ref. \cite{zhu-prd}. The recoupling coefficients are listed in
the Appendix.

\section{Numerical Results}\label{sec3}

In order to estimate the heavy tetraquark mass, we need determine
the constituent quark mass and the value of the parameter
$\bar{v}$ in Eq. (\ref{v-Q}) for the heavy tetraquark system.
Using the experimental values of $J/\psi, \eta_{c}, D^{*}, D,
D_{s}^{*}, D_{s}, \Upsilon$ masses, we use Eq. (\ref{p1}) to
extract the constituent quark mass:
\begin{eqnarray}
\begin{cases}
M(J/\psi)=2m_c+\frac{16}{3}v_{c \bar c}(\frac{m_{u}}{m_{c}})^{2} & \\
M(\eta_{c})=2m_c-16v_{c\bar c}(\frac{m_{u}}{m_{c}})^{2}&\\
M(D^{*})=m_u+m_c+\frac{16}{3}v_{c \bar u}(\frac{m_{u}}{m_{c}})&\\
M(D)=m_u+m_c-16v_{c \bar u}(\frac{m_{u}}{m_{c}})&\\
M(D_{s}^{*})=m_s+m_c+\frac{16}{3}v_{c \bar s}(\frac{m_{u}}{m_{c}})(\frac{m_{u}}{m_{s}})&\\
M(D_{s})=m_s+m_c-16v_{c \bar s}(\frac{m_{u}}{m_{c}})(\frac{m_{u}}{m_{s}})&\\
M(\Upsilon)=2m_b+\frac{16}{3}v_{b\bar b}(\frac{m_{u}}{m_{b}})^{2} & \\
\end{cases}
\Rightarrow\begin{cases}m_c = 1534 \mbox{MeV} \\
m_u = 437  \mbox{MeV} \\
  m_s = 542 \mbox{MeV} \\
  m_b = 4730 \mbox{MeV} \\
v_{c \bar u}=23.4 \mbox{MeV}\\ v_{c \bar s}=29.3
\mbox{MeV} \\
v_{c \bar c} = 67.7 \mbox{MeV}
\end{cases}
\end{eqnarray}
where $ v_{c \bar u}$, $v_{c \bar s}$ and $v_{c \bar c}$ are
merely the CM interaction parameters in Eq. (\ref{hcm}) for the
$c\bar{c},c\bar{u},c\bar{s}$ meson system respectively.

With the assumption that X(3872) is a $qc\bar q\bar c$ tetraquark,
the $V_{CM}(qc\bar q\bar c)$ reads
\begin{equation}\label{HCORcc}
V_{CM}(qc\bar q\bar c)=\zeta_{c}V_{12} +
V_{13}+\zeta_{c} V_{14}
 +\zeta_{c}
V_{23}+ \zeta_{c}^{2}V_{24} +\zeta_{c} V_{34}\;
\end{equation}
where $\zeta_{c}=\frac{m_{u}}{m_{c}}$. After diagonalizing the full
matrix $\langle k \mid V_{CM}(qc\bar q\bar c) \mid l\rangle$, we
obtain CM interaction eigenvalue and mass. The lowest eigenvalue is
\[
V_{CM}=-15.9v_{Q} \; ,
\]
where $v_Q = \bar{v}_{qc\bar q\bar c}$. In the heavy quark limit
$\zeta_{c}=\frac{m_{u}}{m_{c}}\to 0$, the CM energy of $qc\bar
q\bar c$ system is dominated by the $q \bar q$ subsystem
$V_{CM}(qc\bar q\bar c) \to V_{13} = -16v_Q$. The wave function of
this lowest-lying tetraquark state will be $|c\bar c
35,1_{c},1,2\rangle\otimes| q \bar q 1,1_{c},0,2 \rangle$ with
charge conjugate parity $C=-$. This state is strongly coupled to
$J/\psi$ and a light pesudoscalar. Therefore it fall parts very
easily and has a very broad decay width. Thus it is unlikely to
observe it experimentally.  The second lowest-lying tetraquark
state will be $|c\bar c 35,8_{c},1,2\rangle\otimes| q \bar q
35,8_{c},1,2 \rangle$ with $C=+$. We identify it to the observed
X(3872) state:
\begin{eqnarray*}
V_{CM}(X(3872))&=&-4.1v_{Q} \; ,\\ M(X(3872))&=&2m_c +2
m_u-4.1v_{Q} \; .
\end{eqnarray*}
In the heavy quark limit
$\zeta_{c}=\frac{m_{u}}{m_{c}}\to 0$, $V_{CM}(qc\bar q\bar c) \to
V_{13} = -4v_Q$. Using the experimental
value of X(3872) mass as input, we get
\begin{equation}
\nonumber v_{Q}\approx 17.1 \mbox{MeV} \;.
\end{equation}
We will use this $v_Q$ to estimate the mass of other $qq \bar Q \bar Q$
and $qQ \bar q \bar Q$ tetraquarks.

The CM strength parameter ($v$) depends on the constituent quark
mass and spatial wave function of quark model. Since we consider
only the $0^{+}$, $1^{+}$ and $2^{+}$ heavy tetraquarks without
radial and orbital excitations, their spatial wave functions are
expected to be roughly the same for $qc\bar{q}\bar{c}$ and
$qq\bar{c}\bar{c}$ systems. Hence, we use the same parameter $v_Q$
to estimate the CM interaction of $0^{+}$, $1^{+}$ and $2^{+}$
states of all $qQ\bar q \bar Q$ and $qq\bar Q \bar Q$ tetraquark
systems with $Q=c$ or $b$. We also use the same $v_Q$ to estimate
the $qQ\bar Q\bar Q$ heavy tetraquarks. The masses are collected
in Table \ref{tab1}--\ref{tab3}. As pointed above, the
lowest-lying tetraquark state of each $qQ\bar q\bar Q$ system will
be $|Q\bar Q 35,1_{c},1,2\rangle\otimes| q \bar q 1,1_{c},0,2
\rangle$ in the heavy quark limit, which will be too broad to be
observed experimentally. Thus in Table \ref{tab1} we only list the
masses of the second lowest-lying tetraquark states. The neutral
ones of these states have positive C-parity.
\begin{table}[htp]
\begin{ruledtabular}
\begin{tabular}{c d d d d d d}
flavor &\multicolumn{1}{c}{$(0^{+})V_{CM}(v_{Q})$}
&\multicolumn{1}{c}{$(0^{+})M$(MeV)}
&\multicolumn{1}{c}{$(1^{+})V_{CM}(v_{Q})$}
&\multicolumn{1}{c}{$(1^{+})M$(MeV)}
&\multicolumn{1}{c}{$(2^{+})V_{CM}(v_{Q})$}
&\multicolumn{1}{c}{$(2^{+})M$(MeV)}\\
  \hline
$qc\bar q\bar c$&-7.4& 3816&-4.1& 3872 &2.7& 3988    \\

$qc\bar s\bar c$&-6.5& 3936&-3.8&3982&2.5& 4090\\

$sc\bar s\bar c$&-5.6 & 4056&-3.2&4097&2.3& 4191\\

$qb\bar q\bar b$&-3.0& 10283&-1.8& 10303 &0.4& 10341    \\

$qb\bar s\bar b$&-2.6& 10395&-1.6&10412&0.5& 10448\\

$sb\bar s\bar b$&-2.3 &10505 &-1.3&10522&0.5&10553 \\

$qc\bar q\bar b$&-5.4& 7046&-3.9& 7071&1.6& 7165    \\

$qc\bar s\bar b$&-4.5& 7166&-3.1&7190&1.4& 7267\\

$sc\bar s\bar b$&-4.1 &7278 &-3.0&7297&1.4&7371 \\
\end{tabular}
\end{ruledtabular}
\caption{The CM energy and masses of $0^{+}$, $1^{+}$ and $2^{+}$ of
$qQ \bar q \bar Q$. The CM energy is in unit $v_{Q}$. }\label{tab1}
\end{table}

\begin{table}[htp]
\begin{ruledtabular}
\begin{tabular}{c d d d d d d}
flavor &\multicolumn{1}{c}{$(0^{+})V_{CM}(v_{Q})$}
&\multicolumn{1}{c}{$(0^{+})M$(MeV)}
&\multicolumn{1}{c}{$(1^{+})V_{CM}(v_{Q})$}
&\multicolumn{1}{c}{$(1^{+})M$(MeV)}
&\multicolumn{1}{c}{$(2^{+})V_{CM}(v_{Q})$}
&\multicolumn{1}{c}{$(2^{+})M$(MeV)}\\
  \hline
$\{q q\}\bar c\bar c$&-3.9& 3875&1.4& 3966 &4.4& 4017    \\

$\{s s\}\bar c\bar c$&-3.6& 4090&0.7&4164&3.2& 4207\\

$[q q]\bar c\bar c$& & &-9.1&3786&& \\

$q s\bar c\bar c$&-3.8&3982&-7.5&3919&3.7& 4110
\\

$\{q q\}\bar b\bar b$&0.7& 10346&2.2& 10372 &3.2& 10389   \\

$\{s s\}\bar b\bar b$&0.1&10546&1.4& 10568&2.2& 10582\\

$[q q]\bar b\bar b$&&&-8.1&10195&& \\

$u s\bar b\bar b$&0.4& 10446&-6.6&
10326&2.6& 10483\\

$\{q q\}\bar c\bar b$&-1.6& 7111&-0.2& 7135&3.7& 7201   \\

$\{s s\}\bar c\bar b$&-1.8& 7317&-0.7&7336&2.6& 7392\\

$[q q]\bar c\bar b$&-11.1& 6948&-9.3&6979&1.1& 7157\\

$u s\bar c\bar b$&-9.5& 7081&-7.8&
7110&1.2& 7264\\
\end{tabular}
\end{ruledtabular}
\caption{The CM energy and masses of $0^{+}$, $1^{+}$ and $2^{+}$ of
$qq \bar Q\bar Q$. The CM energy is in unit $v_{Q}$.
 $[q q]$ is the
antisymmetrical representation in $SU(3)_{f}$, $\{q q\}$ is the
 symmetrical representation.}\label{tab2}
\end{table}

\begin{table}[htp]
\begin{ruledtabular}
\begin{tabular}{c d d d d d d}
flavor &\multicolumn{1}{c}{$(0^{+})V_{CM}(v_{Q})$}
&\multicolumn{1}{c}{$(0^{+})M$(MeV)}
&\multicolumn{1}{c}{$(1^{+})V_{CM}(v_{Q})$}
&\multicolumn{1}{c}{$(1^{+})M$(MeV)}
&\multicolumn{1}{c}{$(2^{+})V_{CM}(v_{Q})$}
&\multicolumn{1}{c}{$(2^{+})M$(MeV)}\\
  \hline
$qc\bar c\bar c$&-3.5& 4979&-4.0& 4971 &2.0& 5073   \\

$sc\bar c\bar c$&-3.0& 5093&-3.1&5091&1.7& 5173\\

$qb\bar c\bar c$&-3.4 &8177 &-3.4&8177&1.3& 8257\\

$sb\bar c\bar c$&-2.8 &8292 &-2.7&8294&1.1&8359 \\

$qb\bar b\bar b$&-1.0& 14610&-1.4& 14603 &0.5& 14636    \\

$sb\bar b\bar b$&-0.8& 14718&-1.1&14713&0.4& 14739\\

$qc\bar b\bar b$&-0.6 &11421 &-2.5&11388&1.1& 11450\\

$sc\bar b\bar b$&-0.6&11526 &-2.0&11502&0.9&11551 \\

$qc\bar c\bar b$&-5.8& 8136&-4.8& 8153 &0.9& 8250   \\

$sc\bar c\bar b$&-4.8& 8258&-3.9&8273&0.8& 8354\\

$qb\bar c\bar b$&-4.9 &11347 &-4.6&11352&0.5& 11440\\

$sb\bar c\bar b$&-4.0 & 11468&-3.7&11473&0.4&11543 \\
\end{tabular}
\end{ruledtabular}
\caption{The CM energy and masses of $0^{+}$, $1^{+}$ and $2^{+}$ of
$qQ \bar Q\bar Q$. The CM energy is in unit $v_{Q}$.
 $[q q]$ is
antisymmetrical representation in $SU(3)_{f}$, $\{q q\}$ is
 symmetrical representation.}\label{tab3}
\end{table}

\section{Discussions}\label{sec4}

In short summary, we have performed a schematic study of the
masses of possible heavy tetraquarks using the color-magnetic
interaction with the flavor symmetry breaking corrections.
Treating X(3872) as a $qc \bar q \bar c$ tetraquark and using its
mass as input, we extract the CM interaction parameter $v_{Q}$ for
the heavy tetraquark system. With the same $v_Q$, we have
estimated other heavy tetraquark masses. It's interesting to note
from the above tables that the CM interaction is repulsive for
$2^+$ heavy tetraquarks. From Table I, it's clear that $0^+$ $qc
\bar q \bar c$ states will also exist if X(3872) is really a $1^+$
tetraquark.

\section{Acknowledgments}

This project was supported by the National Natural Science Foundation
of China under Grants 10375003 and 10421503, Key Grant Project of
Chinese Ministry of Education (NO 305001). C.Y.
thanks Y. R. Liu and W. Wei for helpful discussions.

\appendix

\section{The recoupling coefficients}

In order to calculate the terms of $V(q \bar q)$, we need do the following
recouplings \cite{belle9, belle10}£º
\begin{eqnarray}
\{|q_{1}q_{2}D_{6}(Q)D_{3}(Q)S(Q)\rangle\otimes| \bar q_{3}\bar q_{4}D_{6}(\bar
Q)D_{3}(\bar Q)S(\bar Q)\rangle\}_{(D_{3},S)}\nonumber\\
=\sum R(D_{3}(Q)D_{3}(\bar Q);D_{3}(13)D_{3}(24);D)R(S(Q)S(\bar
Q);S(13)S(24);S)\nonumber\\
\times\{|q_{1}\bar q_{3}D_{6}(13)D_{3}(13)S(13)\rangle\otimes| q_{2}\bar
q_{4}D_{6}(24)D_{3}(24)S(24)\rangle\}_{(D_{3},S)}\;,
\end{eqnarray}
where the recoupling coeffiecients are:
\begin{eqnarray}
&&R(S(Q)S(\bar Q);S(13)S(24);S)\nonumber\\
&=&\sqrt{(2S(Q)+1)(2S(\bar
Q)+1)(2S(13)+1)(2S(24)+1)}\bordermatrix{&&&\cr
&\frac{1}{2}&\frac{1}{2}&S(Q)\cr &\frac{1}{2}&\frac{1}{2}&S(\bar
Q) \cr &S(13)&S(24)&S \cr}
\;,\\
&&R((\lambda_{Q}\mu_{Q})(\lambda_{\bar Q}\mu_{\bar
Q});(\lambda_{13}\mu_{13})(\lambda_{24}\mu_{24}))\nonumber\\
&=&(-1)^{\lambda_{Q}+\mu_{Q}+\lambda_{13}+\mu_{13}}U((10)(10)(10)(01);
(\lambda_{Q}\mu_{Q})(\lambda_{13}\mu_{13}))\;.
\end{eqnarray}

We will list the recoupling coefficients of $0^{+}$, $1^{+}$ and
$2^{+}$ heavy tetraquarks below. With $q_{1}q_{2} \otimes \bar
q_{3} \bar q_{4} $ construction, the $0^{+}$ heavy tetraquarks
are:
\begin{eqnarray}\label{en00}
 |1,1_{c},
S=0,4\rangle&=&\sqrt{\frac{6}{7}}|21,6_{c},1,2;
\overline{21},\overline{6}_{c},1,2\rangle
+\sqrt{\frac{1}{7}}|21,\overline{3}_{c},0,2; \overline{21},3_{c},0,2\rangle \nonumber\\
|405,1_{c},0,4\rangle&=&\sqrt{\frac{1}{7}}|21,6_{c},1,2;
\overline{21},\overline{6}_{c},1,2\rangle
-\sqrt{\frac{6}{7}}|21,\overline{3}_{c},0,2; \overline{21},3_{c},0,2\rangle\nonumber \\
|1,1_{c},0,4\rangle&=&\sqrt{\frac{3}{5}}|15, \overline{3}_{c},1,2;
\overline{15},3,1,2\rangle
+\sqrt{\frac{2}{5}}|15,6_{c},0,2; \overline{15},\overline{6}_{c},0,2\rangle \nonumber\\
|189,1_{c},0,4\rangle&=&\sqrt{\frac{2}{5}}|15, \overline{3}_{c},1,2;
\overline{15},3,1,2,\rangle -\sqrt{\frac{3}{5}}|15,6_{c},0,2;
\overline{15},\overline{6}_{c},0,2\rangle
\end{eqnarray}
For $0^{+}$ tetraquarks, the basis states of the
$q_{1}q_{2}\otimes \bar q_{3} \bar q_{4}$ construction are:
\begin{eqnarray}\label{en0a}
a^{0}_{1}&=&|q_{1}q_{2} 21,6_{c},S=1;\bar q_{3} \bar q_{4}
\overline{21},\overline{6}_{c},1\rangle;
a^{0}_{2}=|q_{1}q_{2} 21,\overline{3}_{c},0;\bar q_{3} \bar q_{4} \overline{21},3_{c},0\rangle \nonumber\\
a^{0}_{3}&=&|q_{1}q_{2} 15, \overline{3}_{c},1;\bar q_{3} \bar q_{4}
\overline{15},3,1\rangle; \ \ \ \ \ \ \  a^{0}_{4}=|q_{1}q_{2}
15,6_{c},0;\bar q_{3} \bar q_{4}
\overline{15},\overline{6}_{c},0\rangle
\end{eqnarray}
The corresponding basis states of the $q_{1} \bar q_{3} \otimes
q_{2} \bar q_{4}$ construction are:
\begin{eqnarray}\label{en0b}
b^{0}_{1}&=&|q_1\bar q_{3} 1,1,0;  q_2 \bar q_{4} 1,1,0
 \rangle; \ \ \
b^{0}_{2}=|q_1\bar q_{3} 35,1,1;  q_2 \bar q_{4} 35,1,1
 \rangle \nonumber\\
b^{0}_{3}&=&|q_1\bar q_{3} 35,8,0; q_2 \bar q_{4} 35,8,0\rangle;
b^{0}_{4}=||q_1\bar q_{3} 35,8,1; q_2 \bar q_{4}
35,8,1\rangle
\end{eqnarray}
The transform matrix from Eq. (\ref{en0a}) to Eq. (\ref{en0b}) is:
\begin{equation}
\bordermatrix{&&&&\cr
&\frac{\sqrt{2}}{2}&-\frac{\sqrt{6}}{6}
&\frac{1}{2}&-\frac{\sqrt{3}}{6}\cr
 &\frac{\sqrt{3}}{6}&\frac{1}{2}
&-\frac{\sqrt{6}}{6}&-\frac{\sqrt{2}}{2}\cr &\frac{1}{2}
&-\frac{\sqrt{3}}{6}&-\frac{\sqrt{2}}{2}&\frac{\sqrt{6}}{6}\cr
&\frac{\sqrt{6}}{6}&\frac{\sqrt{2}}{2}
&\frac{\sqrt{3}}{6}&\frac{1}{2}\cr}
\end{equation}
Then we can write the $0^{+}$ $SU(6)$ eigenstates in terms of two
pairs of $q_1\bar q_3 \otimes q_2\bar q_4$:
\begin{eqnarray}\label{en01}
|1,1_{c},0\rangle
 &=&\frac{\sqrt{21}}{6}|q_1\bar q_{3}
1,1,0;  q_2 \bar q_{4} 1,1,0
 \rangle
 -\frac{\sqrt{7}}{14}|q_1\bar q_{3}
35,1,1;  q_2 \bar q_{4} 35,1,1
 \rangle\nonumber\\
&+&\frac{\sqrt{42}}{21}|q_1\bar q_{3} 35,8,0; q_2 \bar q_{4}
35,8,0\rangle
 -\frac{\sqrt{14}}{7}|q_1\bar q_{3} 35,8,1; q_2
\bar q_{4} 35,8,1
 \rangle\\
|405,1_{c},0\rangle
 &=&-\frac{2\sqrt{42}}{21}|q_1\bar q_{3} 35,1,1; q_2
\bar q_{4} 35,1,1
 \rangle
 +\frac{3\sqrt{7}}{14}|q_1\bar q_{3} 35,8,0; q_2
\bar q_{4} 35,8,0\rangle \nonumber\\
&&  +\frac{5\sqrt{21}}{42}|q_1\bar q_{3} 35,8,1; q_2 \bar q_{4}
35,8,1  \rangle\\
 |1,1_{c},0\rangle
 &=&\frac{\sqrt{15}}{6}|q_1\bar q_{3}
1,1,0;  q_2 \bar q_{4} 1,1,0
 \rangle
+\frac{\sqrt{5}}{10}|q_1\bar q_{3} 35,1,1;  q_2 \bar q_{4} 35,1,1
 \rangle\nonumber\\
&-&\frac{\sqrt{30}}{15}|q_1\bar q_{3} 35,8,0; q_2 \bar q_{4}
35,8,0\rangle
 +\frac{\sqrt{10}}{5}|q_1\bar q_{3} 35,8,1; q_2
\bar q_{4} 35,8,1
 \rangle\\
|189,1_{c},0\rangle
 &=&-\frac{2\sqrt{30}}{15}|q_1\bar q_{3} 35,1,1; q_2
\bar q_{4} 35,1,1
 \rangle
-\frac{3\sqrt{5}}{10}|q_1\bar q_{3} 35,8,0; q_2
\bar q_{4} 35,8,0\rangle \nonumber\\
&&  -\frac{\sqrt{15}}{30}|q_1\bar q_{3} 35,8,1; q_2 \bar q_{4}
35,8,1 \rangle
\end{eqnarray}

For the $1^{+}$ heavy tetraquarks, the states with the $q_{1}q_{2}
\otimes \bar q_{3} \bar q_{4} $ construction are:
\begin{eqnarray}
\label{11} |35,1_{c},1,4,\rangle &=&
|21,6_{c},1,2,\rangle\otimes|\overline{21},\overline{6}_{c},1,2,\rangle
\nonumber\\
|35,1_{c},1,4\rangle &=&
|15,\overline{3}_{c},1,2,6_{f}\rangle\otimes|\overline{15},3_{c},1,2\rangle
\nonumber\\
|35,1_{c},1,4\rangle &=&
\sqrt{\frac{1}{3}}|21,\overline{3}_{c},0,2\rangle
\otimes|\overline{15},3_{c},1,2\rangle
-\sqrt{\frac{2}{3}}|21,6_{c},1,2\rangle
\otimes|\overline{15},\overline{6}_{c},0,2\rangle
\nonumber\\
|280,1_{c},1,4\rangle &=&
\sqrt{\frac{2}{3}}|21,\overline{3}_{c},0,2\rangle
\otimes|\overline{15},3_{c},1,2,\rangle
+\sqrt{\frac{1}{3}}|21,6_{c},1,2,\rangle
\otimes|\overline{15},\overline{6}_{c},0,2,\rangle
\nonumber\\
|35,1_{c},1,4\rangle &=& \sqrt{\frac{1}{3}}
|15,\overline{3}_{c},1,2\rangle \otimes
|\overline{21},3_{c},0,2\rangle -\sqrt{\frac{2}{3}}
|15,6_{c},0,2\rangle \otimes
|\overline{21},\overline{6}_{c},1,2\rangle
\nonumber\\
 |280,1_{c},1,4\rangle &=&
\sqrt{\frac{2}{3}} |15,\overline{3}_{c},1,2\rangle \otimes
|\overline{21},3_{c},0,2\rangle +\sqrt{\frac{1}{3}}
|15,6_{c},0,2\rangle \otimes
|\overline{21},\overline{6}_{c},1,2\rangle
\end{eqnarray}
The basis states are:
\begin{eqnarray}\label{en1a}
a^{1}_{1}&=&|q_{1}q_{2} 21,\overline{3}_{c},0;\bar q_{3} \bar q_{4}
\overline{15},3_{c},1\rangle;
a^{1}_{2}=|q_{1}q_{2} 21,6,1;\bar q_{3} \bar q_{4} \overline{15},\overline{6}_{c},0\rangle \nonumber\\
a^{1}_{3}&=&|q_{1}q_{2} 21,6,1;\bar q_{3} \bar q_{4}
\overline{21},\overline{6}_{c},1\rangle;\
 a^{1}_{4}=|q_{1}q_{2} 15,\overline{3},1;\bar q_{3} \bar q_{4}
\overline{15},3,1\rangle\;\nonumber\\
a^{1}_{5}&=&|q_{1}q_{2} 15,\overline{3},1;\bar q_{3} \bar q_{4}
\overline{21},3,0\rangle;\ \ a^{1}_{6}=|q_{1}q_{2}
15,6,0;\bar q_{3} \bar q_{4} \overline{21},\overline{6},1\rangle
\end{eqnarray}
With $q_{1} \bar q_{3} \otimes q_{2} \bar q_{4}$ construction, the
basis states are:
\begin{eqnarray}\label{en1b}
b^{1}_{1}&=&|q_1\bar q_{3} 1,1,0;  q_2 \bar q_{4} 35,1,1
 \rangle;\
b^{1}_{2}=|q_1\bar q_{3} 35,8,0;  q_2 \bar q_{4} 35,8,1
 \rangle \nonumber\\
b^{1}_{3}&=&|q_1\bar q_{3} 35,1,1; q_2 \bar q_{4} 1,1,0\rangle;\
b^{1}_{4}=||q_1\bar q_{3} 35,8,1; q_2 \bar q_{4}
35,8,0\rangle\;\nonumber\\
b^{1}_{5}&=&|q_1\bar q_{3} 35,8,1;  q_2 \bar q_{4} 35,8,1
 \rangle;
 b^{1}_{6}=|q_1\bar q_{3} 35,1,1;  q_2 \bar q_{4} 35,1,1
 \rangle
\end{eqnarray}
The transform matrix from Eq. (\ref{en1a}) to Eq. (\ref{en1b}) is:
\begin{equation}
\bordermatrix{&&&&&&\cr &\frac{\sqrt{3}}{6}& -\frac{\sqrt{6}}{6}
&-\frac{\sqrt{3}}{6}&
\frac{\sqrt{6}}{6}&-\frac{\sqrt{3}}{3}&\frac{\sqrt{6}}{6}\cr
 &-\frac{\sqrt{6}}{6}&-\frac{\sqrt{3}}{6}
&\frac{\sqrt{6}}{6}&\frac{\sqrt{3}}{6}&\frac{\sqrt{6}}{6}&\frac{\sqrt{3}}{3}\cr
&\frac{\sqrt{3}}{3}
&\frac{\sqrt{6}}{6}&\frac{\sqrt{3}}{3}&\frac{\sqrt{6}}{6}&0&0\cr
&\frac{\sqrt{6}}{6}&-\frac{\sqrt{3}}{3}
&\frac{\sqrt{6}}{6}&-\frac{\sqrt{3}}{3}&0&0\cr
&\frac{\sqrt{3}}{6}&-\frac{\sqrt{6}}{6}
&-\frac{\sqrt{3}}{6}&\frac{\sqrt{6}}{6}&-\frac{\sqrt{3}}{3}&\frac{\sqrt{6}}{6}\cr
&-\frac{\sqrt{6}}{6}&-\frac{\sqrt{3}}{6}
&\frac{\sqrt{6}}{6}&\frac{\sqrt{3}}{6}&\frac{\sqrt{6}}{6}&\frac{\sqrt{3}}{3}\cr}
\end{equation}
Then the $1^{+}$ $SU(6)$ flavor eigenstates in terms of two pairs
of $q_1\bar q_3 \otimes q_2\bar q_4$ read
\begin{eqnarray}
\label{en1-11} |35,1_c,1
 \rangle
 &=&\frac{1}{2}|q_1\bar q_{3}
1,1,0;  q_2 \bar q_{4} 35,1,1
 \rangle
 -\frac{1}{2}|q_1\bar q_{3}
35,1,1;  q_2 \bar q_{4} 1,1,0
 \rangle\nonumber\\
&-&\frac{2}{3}|q_1\bar q_{3} 35,8,1; q_2 \bar q_{4} 35,8,1
 \rangle
 -\frac{\sqrt{2}}{6}|q_1\bar q_{3} 35,1,1; q_2
\bar q_{4} 35,1,1
 \rangle \\
\label{en1-22} |280, 1_c,1
 \rangle
 &=&-\frac{1}{2}|q_1\bar q_{3} 35,8,0; q_2
\bar q_{4} 35,8,1
 \rangle
 +\frac{1}{2}|q_1\bar q_{3} 35,8,1; q_2
\bar q_{4} 35,8,0
 \rangle\nonumber\\
&-&\frac{\sqrt{2}}{6}|q_1\bar q_{3} 35,8,1; q_2 \bar q_{4} 35,8,1
 \rangle
 +\frac{2}{3}|q_1\bar q_{3} 35,1,1; q_2
\bar q_{4} 35,1,1
 \rangle\\
\label{en1-33}  |35,1_{c},1\rangle &=&\frac{\sqrt{3}}{3}|q_1\bar
q_{3} 1, 1,0;  q_2 \bar q_{4} 35,1,1
 \rangle
+\frac{\sqrt{6}}{6}|q_1\bar q_{3} 35,8,0; q_2 \bar q_{4} 35,8,1
 \rangle\nonumber\\
&+&\frac{\sqrt{3}}{3}|q_1\bar q_{3} 35,1,1;  q_2 \bar q_{4} 1,1,0
 \rangle
 +\frac{\sqrt{6}}{6}|q_1\bar q_{3} 35,8,1; q_2
\bar q_{4} 35,8,0
 \rangle\\
 \label{en1-44}|35,1_{c},1\rangle
 &=&\frac{\sqrt{6}}{6}|q_1\bar q_{3}
1, 1,0;  q_2 \bar q_{4} 35,1,1
 \rangle
 -\frac{\sqrt{3}}{3}|q_1\bar q_{3} 35,8,0; q_2
\bar q_{4} 35,8,1
 \rangle\nonumber\\
&+&\frac{\sqrt{6}}{6}|q_1\bar q_{3} 35,1,1;  q_2 \bar q_{4} 1,1,0
 \rangle
-\frac{\sqrt{3}}{3}|q_1\bar q_{3} 35,8,1; q_2 \bar q_{4} 35,8,0
 \rangle\\
 \label{en1-55}
|35,1_c,1
 \rangle
 &=&\frac{1}{2}|q_1\bar q_{3}
1,1,0;  q_2 \bar q_{4} 35,1,1
 \rangle
 -\frac{1}{2}|q_1\bar q_{3}
35,1,1;  q_2 \bar q_{4} 1,1,0
 \rangle\nonumber\\
&-&\frac{2}{3}|q_1\bar q_{3} 35,8,1; q_2 \bar q_{4} 35,8,1
 \rangle
 -\frac{\sqrt{2}}{6}|q_1\bar q_{3} 35,1,1; q_2
\bar q_{4} 35,1,1
 \rangle \\
 \label{en1-66} |280, 1_c,1
 \rangle
 &=&-\frac{1}{2}|q_1\bar q_{3} 35,8,0; q_2
\bar q_{4} 35,8,1
 \rangle
 +\frac{1}{2}|q_1\bar q_{3} 35,8,1; q_2
\bar q_{4} 35,8,0
 \rangle\nonumber\\
&-&\frac{\sqrt{2}}{6}|q_1\bar q_{3} 35,8,1; q_2 \bar q_{4} 35,8,1
 \rangle
 +\frac{2}{3}|q_1\bar q_{3} 35,1,1; q_2
\bar q_{4} 35,1,1
 \rangle
\end{eqnarray}

For the $2^{+}$ heavy tetraquarks,
\begin{eqnarray}
&&|405,1_{c},2\rangle \nonumber\\
&=&\{|q_{1}q_{2} 21,6,1; \bar q_{3} \bar q_{4} \overline{21},\bar
6,S=1 \rangle\}\nonumber\\
&=&\sqrt{\frac{1}{3}}|q_1\bar q_{3} 35,8,1; q_2 \bar q_{4} 35,8,1
 \rangle
+\sqrt{\frac{2}{3}}|q_1\bar q_{3} 35,1,1; q_2 \bar q_{4}
35,1,1\rangle \;\\
&&|189,1_{c},2\rangle \nonumber\\
&=&\{|q_{1}q_{2} 15,\bar 3,1; \bar q_{3} \bar q_{4}
\overline{15},3,1 \rangle\}\nonumber\\
&=&\sqrt{\frac{2}{3}}|q_1\bar q_{3} 35,8,1; q_2 \bar q_{4} 35,8,1
 \rangle
-\sqrt{\frac{1}{3}}|q_1\bar q_{3} 35,1,1; q_2 \bar q_{4}
35,1,1\rangle
\end{eqnarray}
With $SU_c(3)$ and $SU_s(2)$ symmetry, similar recoupling
coefficients can also be obtained.

\end{document}